\documentclass[a4paper,10pt]{article}

\usepackage{amssymb,amsmath}
\usepackage{amsfonts}
\usepackage{enumitem}
\usepackage{latexsym}
\usepackage{graphicx}

\usepackage[utf8]{inputenc}  
\usepackage[T1]{fontenc} 

\title{Théorie des champs des contraintes et des déformations en relativité générale et expansion cosmologique.}
\author{Mathieu R. Beau\\
School of Theoretical Physics, Dublin Institute for Advanced Studies,\\ 
10 Burlington Road, Dublin 4, Ireland. \\
Courriel : mathieu.beau.89@gmail.com}

\begin{document}

\maketitle

\begin{abstract}
Dans cet article nous proposons de modifier les équations d'Einstein en y ajoutant un tenseur de contraintes générale, qui s'exprime en fonction du champ des déformations. On part du principe que la matière et l'espace-temps métrique qui la contient, constituent un milieu continu qui possède des propriétés élastiques de déformation. Nous donnons alors l'expression du tenseur des contraintes pour le \textit{milieu cosmologique} que l'on suppose spatialement homogène et isotrope, puis dérivons les équations de Friedmann modifiées par la présence du champ des déformations. En première approximation nous retrouvons le terme cosmologique $\Lambda g_{\mu\nu}$ où la constante cosmologique s'exprime comme $\Lambda=K\varepsilon$ ou $K$ s'interprète comme le module d'élasticité isostatique et $\varepsilon$ comme les variations relatives du volume après déformation. Enfin, nous déterminons en seconde approximation des termes correctifs aux solutions prédites par le modèle standard qui dépendent de ces deux nouveaux 
paramètres. 
  
\end{abstract}

\section{Introduction}

La théorie de la gravitation relativiste, formulée par Einstein en 1916 \cite{Einstein1916} dans sa version finale, a prédit depuis de façon surprenante des phénomènes dont nombreux ont étés vérifiées et aucune observation n'a remis en cause la validité des equations. Il y a eu dans l'Histoire de cette théorie qui ne s'est bien entendu pas arrêté à sa génèse, de nombreux débats philosophiques et scientifique sur la meilleure façon de formuler la théorie et sur son interprétation. La référence \cite{aether} donne un bon exemple de débats vifs et passionnés sur le sujet. Pour autant, il n'en reste pas moins que le point de vue développé par Einstein n'a pas été contredit par une observation qui remettrait aussi en cause profondément les paradigmes de la théorie. L'une des plus récente observation qui a probablement été celle qui a surpris le plus la communauté scientifique de l'époque fut la découverte de l'accélération de l'expansion cosmologique \cite{ExpansionAcc}. L'introduction d'une nouvelle énergie fut 
alors nécessaire pour rendre compte de ce phénomène, et on l'a alors appelée \textit{énergie noire} car aucune observation traditionelle basée sur l'optique de cette mystérieuse énergie n'a été faite. Cependant, l'ajout du terme cosmologique $\Lambda g_{\mu\nu}$ dans les équations d'Einstein reste la façon la plus élégante, la plus simple et la plus adéquate de décrire ce phénomène. L'équation d'Einstein s'écrit alors: 
 \begin{equation}\label{Einstein}
 R_{\mu\nu}-\frac{1}{2}R g_{\mu\nu}+\Lambda g_{\mu\nu}=\frac{8\pi G}{c^4}T_{\mu\nu}
 \end{equation}
où se trouve  gaucheà du terme cosmologique le tenseur d'Einstein et où $T_{\mu\nu}$ est le tenseur impulsion-énergie décrivant la matière et le rayonnement présent dans l'Univers. Plusieurs observations \cite{CMB} corroborent l'hypothèse d'une énergie noire ou \textit{énergie du vide} qui représenterait environ $70 \%$ de la quantité totale d'énergie dans l'Univers. L'idée est la suivante, son tenseur impulsion-énergie $T_{\mu\nu}^{(\Lambda)}=\rho_{\Lambda} g_{\mu\nu}$ avec $\rho_\Lambda=\frac{\Lambda c^2}{8\pi G}$ décrit une énergie qui excerce une pression négative $p_\Lambda=-\rho_\Lambda c^2$ sur le volume et aurait donc tendance à 'pousser' la matière. On pourrait donc très bien se contenter de cette description, en assumant que le vide contient une énergie qui pousse la matière (pression négative) provoquant l'accélération de l'expansion. Cependant, différentes questions restent posées aujourd'hui:
\begin{enumerate}
 \item Tout d'abord, il est normal que cette l'hypothèse ad hoc concernant l'ajout d'un terme $\Lambda g_{\mu\nu}$ puisse choquer car on ne connait pas la nature de cette énergie du vide. Correspond-elle à des propriétés physiques de l'espace-temps ? 
 \item Ensuite, on peut se poser la question de savoir pourquoi $\Lambda$ est une constante, c'est à dire invariante dans le temps ? Et si $\Lambda$ dépend du temps, comment intérpréter le fait que l'impulsion-énergie de la matière ne se conserve pas, d'après \eqref{Einstein} en modifiant $\Lambda$ par $\Lambda(t)$. Enfin, comment formuler une théorie consistante qui donne un sens à cette fonction $\Lambda(t)$ ?     
 \item Pour compléter les deux premiers points, il est intéressant de rajouter que si l'espace est vide alors cette énergie provoque un expansion constante (non accélérée), voir un récent article dans le même journal \cite{deBroglie}. Mais du point de vue de Mach-Einstein, cela signifie que l'espace-temps existe sans matière et ainsi la conception d'énergie du vide ne semble pas satisfaisante de ce point de vue qui n'est pas contredit par les observations actuelle. 
\end{enumerate}

Nous arrivons donc à un problème à la fois théorique et empirque. Comment, en gardant le point de vue Mach-Einstein, donner une structure à $\Lambda$ (et donc une explication à l'accélération de l'expansion observée) et en donner une interprétation ? 
Si nous arrivons à une description qui satisfait les observations actuelles et proche du modèle standard qui relate de façon précise ces observations, pouvons-nous donner des corrections et/ou des ajustements aux prédictions théoriques actuelles ?  

Nous proposons ici dans cet article de construire une théorie qui modifie les équations d'Einstein en y ajoutant un terme qui représente les contraintes s'excerçant sur le milieu continu que constituerait la matière-énergie dans l'espace. Nous voulons réinterpréter la constante cosmologique comme un effet de contraintes mécaniques (analogue au module d'élasticité isotrope intervenant dans la théorie de déformation de milieux continus) dues aux déformations de l'espace-temps à l'échelle cosmologique. 
Pour ce faire, nous partons d'abord d'une série de principes que nous énonçons ci-dessous:
\begin{enumerate}[label=(P\arabic*)]
 \item La matière-énergie et l'espace-temps constituent un milieu continu et possède des propriétés élastique de déformation. 
 \item Les contraintes dues aux déformations du milieu modifient l'énergie interne du milieu.
 \item Le champ de déformation génère des courants d'accélérations pour toutes les formes de matière-énergie. De façon équivalente, les trajectoires géodésiques sont modifiées (ou déformées) par ce champ en suivant les lois de déformations du milieu continu.  
\end{enumerate}

Nous allons dans la section suivante établir les équations qui découlent de ces principes. Ensuite dans la section qui suivra, nous dériverons les équations de Friedmann modifiées par la présence des champs de déformation et discuterons les conséquences sur l'évolution cosmologique. 

\section{Construction des équations} 

\indent Commençons ici par établir les équations d'Einstein modifiées par la présence de nouveaux champs que nous allons décrire dans le paragraphe qui suit.

 Le principe (P1) suppose d'une part que le milieu cosmologique est continu et qu'il peut se déformer sous l'effet des contraintes internes de façon analogue aux deformations élastiques d'un milieu tri-dimensionnel en mécanique Newtonienne. 
 Ainsi nous devons formuler une théorie de déformation élastique relativiste dans le cas général où s'ajoute un champ métrique. 
 Nous supposons les déformations suffisamment petites pour pouvoir se limiter à la théorie linéaires des déformations. Nous reviendrons dans la dernière section sur cette hypothèse. Nous introduisons donc les tenseurs suivants:
\begin{gather}
\epsilon_{\mu\nu}=\frac{1}{2}\left(D_{\mu}G_\nu+D_{\nu}G_\mu\right) \label{epsilon} \\
\sigma_{\mu\nu}=C_{\mu\nu\delta\gamma}\epsilon^{\delta\gamma} \label{sigma} 
\end{gather}
où $\epsilon_{\mu\nu}$ est le tenseur des déformations qui s'exprime comme le gradient symétrique du vecteur de déformation $G_\mu$, $\sigma_{\mu\nu}$ est le tenseur des contraintes (dimension d'une pression $N.m^{-2}$ ou densité d'énergie par unité de volume $J.m^{-3}$) reliés de façon linéaires au tenseur des déformations via le tenseur d'élasticité $C_{\mu\nu\delta\gamma}$ qui caratérise les contraintes exercées sur le milieu sous l'effet de petites déformations. Dans ce cas relativiste s'ajoute des composantes temporelles: $\epsilon_{00}$ qui peut se voir comme une contraction/dilatation temporelle, $\epsilon_{0i},\ i=x,y,z$ qui s'interprètent comme les cisaillements spatio-temporels similaires aux cisaillements spatiaux $\epsilon_{ij},\ i\neq j$. Les parties diagonales $\epsilon_{ii},\ i=x,y,z$ sont les contractions/dilatations spatiale du volume. Ainsi le tenseur des contraintes possède lui aussi des parties temporelles $\sigma_{00}$ pouvant s'interpréter comme une énergie de déformation quand $\sigma_{
ii},\ i=x,y,z$ sont les pressions s'exerçant sur les surfaces du volumes. A noté que de façon similaire aux déformations tri-dimensionnelles, l'équation \eqref{epsilon} impose des conditions de compatibilités données par ces équations:  
\begin{gather}
\ D_\gamma D^\gamma \epsilon_{\mu\nu}+D_\mu D_\nu \epsilon_{\gamma}^{\gamma} 
=D_\mu D^\gamma \epsilon_{\gamma\nu}+D_\nu D^\gamma \epsilon_{\gamma\mu} \label{compat} 
 \end{gather}
 
 Le principe (P2) généralise la théorie de déformation élastique d'un milieu continu au cas relativiste en présence de gravité. Autrement dit, les contraintes sur le milieu possède une énergie de déformation (que l'on appelle \textit{énergie de déformation élastique}) et donc d'après les équations d'Einstein doivent aussi contribuer au tenseur impulsion-énergie. Il s'ensuit naturellement que les équations d'Einstein modifiées sont données par:
 \begin{equation}
  R_{\mu\nu}-\frac{1}{2}R g_{\mu\nu}
=\frac{8\pi G}{c^4}\left(T_{\mu\nu}+\sigma_{\mu\nu}\right) \label{EinsteinModif}
 \end{equation}
 De cette équation, nous obtenons la conservation totale de l'impulsion-énergie:
 \begin{equation}
 D_\mu \sigma^{\mu\nu}=-D_\mu T^{\mu\nu} \label{EnergyCons} 
 \end{equation}
 Ainsi, le tenseur impulsion-énergie usuel n'est plus conservé (en général). Sa divergence est égale à l'opposé de la divergence du tenseur des contraintes. De façon tout à fait similaire, nous avons ce résultat pour un milieu continu tri-dimensionnel, \cite{MMC}. Nous pouvons avoir une autre lecture de cette équation que nous discuterons dans le prochain paragraphe.
\footnote{On est en droit de se poser la question de la nature de ces champs de déformation $\epsilon_{\mu\nu}$ et $G_\mu$. On rejoint le même problème concernant l'existence du champ de gravité. Existe-t-il vraiment ou n'est-il qu'une illusion d'optique? D'après le point de vue de Mach-Einstein, la matière est l'origine de ce champ de gravité. Sans matière il n'y a pas d'espace et la notion d'espace vide n'a pas de sens. Il en est de même ici, la notion de déformation du milieu n'existe qu'à condition que la matière soit présente dans ce milieu qui lui-même n'existe que sous la même condition. Les deux aspects différents mais complémentaires, géométrie du milieu (tenseur métrique) et déformation du milieu (tenseur des déformations), ne sont que des conséquences de la présence de matière. On conserve le point de vue Mach-Einstein, mais en ajoutant des propriétés d'élasticité au milieu. On peut rapprocher le tenseur des contraintes à l'hypothèse de l'aether, mais qui serait d'une nature différente de l'aether 
tel qu'on le considère habutellement (voir les références de \cite{aether}) puisque dans notre cas il ne sagit pas d'une substance matérielle. La question de l'aether en relativité général à par ailleurs été discuté par Einstein à plusieurs reprises, voir \cite{EinteinEther1}, \cite{EinteinEther2}, \cite{EinteinEther3}.}
 
 Le dernier principe (P3) est le plus difficile à formuler mathématiquement car il nécessite une généralisation des déformations des trajectoires dans un milieu continu quadri-dimensionnel. Tout d'abord commençons par une remarque pour bien comprendre le problème. Imaginons que nous ayons une distribution de matière et notons son tenseur impulsion-énergie $T_{\mu\nu}$. D'après \eqref{EnergyCons}, les équations du mouvement pour cette distribution ne peut pas être les équations géodésiques habituelles. En fait, ces équations sont analogue au premier groupe d'équations de Maxwell pour le champ électromagnétique qui relient la divergence du tenseur de Faraday $F_{\mu\nu}$ au courant de charges $j^\nu$:
 \begin{equation}\label{Maxwell1}
  D_\mu F^{\mu\nu}=j^\nu
 \end{equation}
 Ces dernières équations signifie que d'un courant de charge émerge un champ électromagnétique. Réciproquement, un champ électromagnétique induit un courant de charge:
 \begin{equation}\label{DynamicCharge}
 \frac{D(\rho_m u_\mu)}{Ds}=\mu_0 F_{\mu\nu} j^{\nu}
 \end{equation}
 où $\rho_m u_\mu$ est le courant de masses chargées en mouvement et où $\mu_0$ est la perméabilité magnétique du vide. Cette équation décrit le phénomène d'induction électromagnétique. D'après les équations d'Einstein \eqref{Einstein} on a $D_\mu T^{\mu\nu}_{(charges)}=-D_\mu T^{\mu\nu}_{(EM)}$ et ces équations contiennent les équations du champs éléctromagnétique si l'on assume les équations du mouvement de la distribution de charge \eqref{DynamicCharge}. Pour notre problème, nous avons un comportement analogue du tenseur des contraintes vis-à-vis de la distribution de masses en mouvement accéléré. En effet, nous pouvons lire les équations \eqref{EnergyCons} de la même manière que \eqref{Maxwell1}, à savoir qu'un courant de masses accélérées génère des contraintes sur le milieu. Réciproquement, nous devrions donc établir des équations décrivant le mouvement de la distribution de masses sous l'effet du champ de déformation $\epsilon_{\mu\nu}$, similaire au principe d'induction électromagnétique formalisé 
par les équations \eqref{DynamicCharge}. En d'autres termes, de la même façon que $A_\mu$ couple avec le courant de charge $j^\mu$:
 $$\Lambda_{\mathrm{Coupl.}}=A_\mu j^\mu$$
 le couplage entre le courant de masses accélérées et le champ de déformation devrait s'écrire:
 \begin{equation}\label{couplage}
 \Lambda_{\mathrm{Coupl.}}=-G_\mu\ D_{\nu}T^{\mu\nu}\ .
 \end{equation}
 où le vecteur de déformation $G_\mu$ induit un courant d'accélération $D_\mu T^{\mu\nu}$ puisque en intégrant par partie ce Lagrangien, on trouve l'équivalent 
 \begin{equation}\label{couplage2}
 \tilde{\Lambda}_{\mathrm{Coupl.}}=\epsilon_{\mu\nu}\ T^{\mu\nu}
 \end{equation}
   Ajouté au Lagrangien libre, le Lagrangien \eqref{couplage2} correspond à une déformation des géodésiques puisque l'on a les transformations suivantes: $$g_{\mu\nu}(x)\mapsto g_{\mu\nu}'(x)=g_{\mu\nu}(x)+\epsilon_{\mu\nu}(x)$$ qui mènent aux équations dynamiques modifiées suivantes:
  \begin{equation}\label{GeodesicModif}
  D_{\mu}T_{\nu}^{\mu}=\partial_{\mu}T_{\nu}^{\mu}+\Gamma^{\mu}_{\mu\gamma}T_{\nu}^{\gamma}-\Gamma^{\gamma}_{\nu\mu}T_{\gamma}^{\mu} 
  =-\epsilon_{\nu}^{\gamma}\partial_{\mu}T_{\gamma}^{\mu}-\Delta^{\mu}_{\mu\gamma}T_{\nu}^{\gamma}+\Delta^{\gamma}_{\nu\mu}T_{\gamma}^{\mu}
  \end{equation}
 où les $\Delta_{\mu\nu\gamma}=\frac{1}{2}\left(\partial_{\nu}\epsilon_{\mu\gamma}+\partial_{\gamma}\epsilon_{\mu\nu}-\partial_{\mu}\epsilon_{\nu\gamma}\right)$ sont analogues aux symboles de Christoffel et correspondent à leur déformation sous l'action de $G_\mu$. 
 \footnote{Il est clair que le champ fondamental $G_\mu$ est vectoriel et donc de spin 1. Les équations \eqref{EnergyCons} et \eqref{compat} nous permettent d'écrire les équations de propagation du champ vectoriel et du tenseur de déformation tout comme on peut le faire pour le potentiel électromagnétique $A_\mu$ et pour le tenseur de Faraday à partir des deux groupes des équations de Maxwell. Ainsi, en supposant l'existence de ce vecteur de déformation, on peut établir les équations d'ondes provoquées par de petites perturbations sur l'accélération de masses en mouvement, ces ondes se propageant dans l'espace. Ceci pourrait être une signature du phénomène de déformation mais on peut imaginer que ces ondes soient très difficiles à détecter. } 
 
 \textit{Remarques.} 
 \begin{itemize}
  \item Une autre façon peut-être plus direct de voir cette effet de déformation géodésique est de regarder l'équation du mouvement pour une particule ponctuelle qui subit le champs de gravitation et le champ de déformation. L'action de la particule est donn\'ee par:
 \begin{equation*}
 S=mc\int ds - \int mc^2 G_\mu(x) \frac{D \dot{x}^\mu}{Ds} ds\ .
 \end{equation*}
 On peut montrer facilement que le deuxi\`eme terme du membre de droite de l'action pr\'ec\'edente 
 est \'equivalent \`a l'action suivante (int\'egration par partie et annulation de l'int\'egrale de la d\'eriv\'ee totale):
 \begin{equation*}
 \int mc^2 \epsilon_{\mu\nu}(x)\dot{x}^{\mu}\dot{x}^{\nu} ds\ ,
 \end{equation*}
 En variant l'action, on obtient les \'equations du mouvement  
 \begin{equation*}
 g_{\mu\nu}\ddot{x}^{\nu}+\Gamma_{\mu\nu\sigma}\dot{x}^{\nu}\dot{x}^{\sigma}= 
 -\left(\varepsilon_{\mu\nu}\ddot{x}^{\nu}+\Delta_{\mu\nu\sigma}\dot{x}^{\nu}\dot{x}^{\sigma}\right)\ ,
 \end{equation*}
 où l'on retrouve les déformations $g_{\mu\nu}'=g_{\mu\nu}+\epsilon_{\mu\nu}$ que nous venons de voir dans le cas général d'une distribution de matière. 
 \item Mentionnons que l'équation \eqref{GeodesicModif} est valable pour toute forme de distribution de matière-énergie (par exemple: tenseur énergie électromagnétique). 
 \item Il faut remarquer que les équations \eqref{GeodesicModif} correspondent à des déformations linéaires (i.e., au premier ordre en $\epsilon_{\mu\nu}$ et $\partial_\sigma\epsilon_{\mu\nu}$) des trajectoires géodésiques. Ici encore nous restreignons notre étude aux petites déformations, mais nous verrons que cela suffit pour notre étude dans la dernière section. Ainsi, en toute généralité nous noterons dans la suite $D_\mu^{(1)}$ l'opérateur définit par:
 \begin{equation}\label{D1}
 D^{(1)}_{\mu}T_{\nu}^{\mu}=\epsilon_{\nu}^{\gamma}\partial_{\mu}T_{\gamma}^{\mu}+\Delta^{\mu}_{\mu\gamma}T_{\nu}^{\gamma}-\Delta^{\gamma}_{\nu\mu}T_{\gamma}^{\mu} 
 \end{equation}
 correspondant à la perturbation de $D_\mu$ au premier ordre. 
 \item Dans le cas où l'impulsion-énergie totale est composée de différentes sortes de champs, par exemple de rayonnement électromagnétique en plus d'une distribution de charges $T_{\mu\nu}=T_{\mu\nu}^{(Charges)}+T_{\mu\nu}^{(EM)}$, le raisonnement est similaire. Il suffit d'écrire les équations du mouvement pour le champs électromagnétique \eqref{Maxwell1} et pour la distribution de charges \eqref{DynamicCharge} puis de déformer les géodésiques en suivant la règle décrite ci-dessus. Notons que dans ce cas les équations \eqref{EnergyCons} combinées avec \eqref{DynamicCharge} ne redonnent plus les équations de Maxwell \eqref{Maxwell1} car le tenseur impulsion-énergie total $T_{\mu\nu}$ de la matière et du champ ne se conserve plus. 
 \end{itemize}

\section{Modèle d'expansion cosmologique}

\subsection{Synthèse: système d'équations}

Nous commençons par rappeler le système d'équation que nous avons établi dans la section précédente et donnons quelques commentaires.

Le système d'équations proposé ci-dessus se résume en trois groupes d'équations:
\begin{gather}
R_{\nu}^{\mu}-\frac{1}{2}R \delta_{\nu}^{\mu}
=\frac{8\pi G}{c^4}\left(T_{\nu}^{\mu}+\sigma_{\nu}^{\mu}\right) \label{E1} \\
D_\mu \sigma_{\nu}^{\mu}=-D_\mu T_{\nu}^{\mu} \label{E2}\\ 
D_{\mu}T_{\nu}^{\mu}=-D_{\mu}^{(1)}T_{\nu}^{\mu}\label{E3} 
\end{gather}
où les quantités nouvelles sont définies ci-dessous:
\begin{gather}
\sigma_{\mu\nu}=C_{\mu\nu\delta\gamma}\epsilon^{\delta\gamma} \label{E4} \\
\ D_\gamma D^\gamma \epsilon_{\mu\nu}+D_\mu D_\nu \epsilon_{\gamma}^{\gamma} 
=D_\mu D^\gamma \epsilon_{\gamma\nu}+D_\nu D^\gamma \epsilon_{\gamma\nu} \label{E5} \\ 
D^{(1)}_{\mu}T_{\nu}^{\mu}=\epsilon_{\nu}^{\gamma}\partial_{\mu}T_{\gamma}^{\mu}+\Delta^{\mu}_{\mu\gamma}T_{\nu}^{\gamma}-\Delta^{\gamma}_{\nu\mu}T_{\gamma}^{\mu} \label{E6} 
\end{gather}
Nous allons commenter ci-dessous ces équations pour rappeler leur signification: 
\begin{itemize}
 \item Les équations d'Einstein modifiées où nous avons ajouté le tenseur des contraintes décrivant les propriétés élastiques du milieu continu, d'après les principes (P1) et (P2). Le tenseur des contraintes défini par \eqref{E4} s'écrit comme une combinaison linéaires des déformations décrites par le tenseur de déformation qui obéit aux conditions de compatibilités \eqref{E5}. Ces conditions sont équivalentes à la définition \eqref{epsilon} et signifient que les déformations s'expriment comme le gradient du vecteur de déformation $G_\mu$.      
 \item Les équations de conservation de l'énergie totale se dérivent des équations \eqref{E1} en prenant la divergence des tenseurs qui donne zero pour le membre de gauche et donc zero pour celui de droite. Elle peut s'interpréter de la façon suivante: l'accélération totale dans l'Univers est nulle, i.e., l'Univers est en inertie de façon globale mais la présence des contraintes sur le milieu induisent une accélération de la matière-énergie. 
 \item Un groupe d'équations supplémentaire est nécessaire pour assurer la consistance du groupe d'équations, c'est à dire qu'en connaissant toutes les conditions initiales le système est entièrement soluble. Les équations \eqref{E3} signifient que la matière couple avec le tenseur de déformation $\epsilon_{\mu\nu}$ de sorte que les équations géodésiques sont modifiées ou déformées par la présence du champ de déformation $\epsilon_{\mu\nu}$. Le perturbation au premier ordre est donnée par \eqref{E6}. Nous appelerons l'ensemble des équations \eqref{E2}-\eqref{E3} les équations d'induction par analogie avec les équations de Maxwell et celles de Lorentz. 
\end{itemize}

\subsection{Tenseur des contraintes et principe cosmologique}

\indent Nous allons discuter le tenseur d'élasticité et proposer une forme adaptée au principe cosmologique. 

De façon général le tenseur d'élasticité est un tenseur compliqué de rang 4. Mais l'espace possède heureusement des propriétés de symétries particulières qui vont nous permettre de réduire le nombre de paramètres et de donner une expression simple et pratique du tenseur des contraintes. Nous allons tout d'abord rappeler le principe cosmologique couramment utilisé et admis, il constitue par ailleurs le quatrième principe que nous utilisons pour la théorie développée ici:
\begin{enumerate}[label=(P4)]
 \item \textit{Principe cosmologique.} L'univers est spatialement homogène et isotrope. 
\end{enumerate}
De ce principe découle directement que le tenseur des contraintes s'écrit:
\begin{equation}\label{sigmaP40}
 \sigma_{\mu\nu}=\alpha\epsilon_{\gamma}^{\gamma}g_{\mu\nu}+2\beta\epsilon_{\mu\nu}
\end{equation}
où $\alpha$ et $\beta$ sont les coefficients de Lamé. On peut réécrire le tenseur sous une autre forme plus pratique pour la suite: 
\begin{equation}\label{sigmaP4}
 \sigma_{\mu\nu}=B\epsilon_{\gamma}^{\gamma}g_{\mu\nu}+2S\left(\epsilon_{\mu\nu}-\frac{1}{3}\epsilon_{\gamma}^{\gamma}g_{\mu\nu}\right)
\end{equation}
où $B=\alpha+2\beta/3$ est le module d'élasticité isostatique et $S=\beta$ est le module de cisaillement, tous deux de la dimension d'une pression. Notons que $\epsilon_{\gamma}^\gamma$ est la trace du tenseur de déformation et caractérise la variation relative du volume sous une pression isostatique pour une petite déformation :
$$\frac{\delta V}{V}\approx\epsilon_\gamma^\gamma$$

On notera dans la suite $\epsilon(t)$ la trace du tenseur des déformation qui ne dépend que de la variable temps $t$ d'après le principe cosmologique. Nous rajoutons une hypothèse à notre modèle: 
\begin{enumerate}[label=(H1)]
 \item \textit{Fluide hydrostatique.} Pour tout temps $t$ fixé, la distribution de matière dans l'espace constitue un fluide hydrostatique en condition d'équilibre, c'est à dire que le volume est soumis à une pression isostatique, sans effets de cisaillement ($S=0$).
\end{enumerate}
 Cette hypothèse est tout à fait acceptable en raison des observations actuelles allant dans ce sens \cite{CMB}. Dans ce cas, le tenseur des contraintes devient simplement:
\begin{equation}\label{sigmaP4B}
 \sigma_{\mu\nu}=B\ g_{\mu\nu}\epsilon(t)\ .
\end{equation}
En prenant $B<0$ et $\epsilon(t)>0$ on retrouve l'interprétation que nous donne le modèle standard, à savoir que l'énergie du vide excerce une pression négative sur le volume. Nous reviendrons sur cette observation dans la suite.

\subsection{Equations de Friedmann modifiées et expansion cosmologique}

Dans cette partie, nous voulons dériver les équations de Friedmann modifiées par l'introduction du champ de déformation et du tenseur des contraintes dans les équations d'Einstein \eqref{E1}. Nous verrons aussi la conséquence importante des équations d'inductions \eqref{E2}-\eqref{E3} dans l'évolution du système cosmologique. Le but que nous nous fixons, comme expliqué dans l'introduction, est de dériver des termes correctifs aux solutions prédites par le modèle standard de la cosmologie. 

Nous considérons ici la métrique de Friedmann-Lemaître-Robertson-Walker (FLRW) en raison du principe cosmologique. Cette métrique décrit l'Univers dans le référentiel des coordonnées comobiles, à savoir $u^\mu=\delta_\mu^0$. Nous rappelons que:
$$ds^2=-c^2dt^2+a(t)^2\left(\frac{dr^2}{1-\kappa r^2}+r^2d\Omega^2\right)$$
où nous avons choisi la signature $(-+++)$, avec en coordonnées polaires $d\Omega^2=d\theta^2+sin(\theta)^2d\phi^2$. Nous notons ici $a(t)$ le facteur d'échelle. Les observations du fond diffus cosmologique \cite{CMB} montrent que $\kappa=0$. Nous prendrons donc cette valeur de $\kappa$ dans la suite. 

Ensuite, nous considérons un gaz non-relativiste qui décrit la matière visible et invisible (dit matière noire). Cette matière est très bien décrite par le tenseur impulsion-énergie d'un gaz parfait de pression nulle:
\begin{equation}\label{T}
T_{\mu\nu}=\rho c^2 u^\mu u^\nu=\rho c^2\delta_{\mu}^{0}\delta_{\nu}^{0} 
\end{equation}
dans les coordonnées comobiles. D'après \eqref{sigmaP4B}, en posant $B=-K$ on a:
\begin{equation}\label{sigmaP4K}
 \sigma_{\mu\nu}=-K\ g_{\mu\nu}\epsilon(t)\ .
\end{equation}
En réécrivant les équations d'Einstein \eqref{E1} de la façon suivante:
$$R_{\mu}^{\nu}=\frac{8\pi G}{c^4}\left(\tilde{T}_{\mu}^\nu-\frac{g_{\mu\nu}}{2}\tilde{T}\right)$$
avec $\tilde{T}_{\mu\nu}=T_{\mu}^{\nu}+\sigma_{\mu}^{\nu}$ and $\tilde{T}=\tilde{T}_\mu^\mu$, on dérive facilement les équations de Friedmann modifiées:
\begin{gather}
3\frac{\ddot{a}}{a}=-\frac{4\pi G}{c^2}\left(\rho(t)c^2+K\epsilon(t)\right) \label{MFE11} \\
\frac{\ddot{a}}{a}+2\frac{\dot{a}^2}{a^2}=\frac{4\pi G}{c^2}\left(\rho(t)c^2+K\epsilon(t)\right)\label{MFE12}
\end{gather}
Ainsi, en combinant les deux dernières équations \eqref{MFE11}-\eqref{MFE12}, nous obtenons:
\begin{equation}\label{MFE1}
\frac{\dot{a}^2}{a^2}=\frac{8\pi G}{3c^2}\left(\rho(t)c^2+K\epsilon(t)\right) 
\end{equation}

Maintenant nous allons dériver l'équation de conservation de l'énergie totale. D'après \eqref{E2} en prenant $\nu=0$ nous avons:
$$\partial_{\mu}T_{0}^{\mu}+\Gamma^{\mu}_{\mu\gamma}T_{0}^{\gamma}-\Gamma^{\gamma}_{0\mu}T_{\gamma}^{\mu}=
\partial_{\mu}\sigma_{0}^{\mu}+\Gamma^{\mu}_{\mu\gamma}\sigma_{0}^{\gamma}-\Gamma^{\gamma}_{0\mu}\sigma_{\gamma}^{\mu}$$
où il est facile de voir que:
\begin{gather*}
\Gamma^{\gamma}_{0\mu} = \frac{1}{c}\frac{\dot{a}}{a},\ \mathrm{si}\ \mu,\gamma\neq0
\end{gather*}
et que donc d'après \eqref{T} et \eqref{sigmaP4K} on a:
\begin{gather*}
\partial_{0}T_{0}^{0}+\frac{3}{c}\frac{\dot{a}}{a}T_{0}^{0}-\frac{1}{c}\frac{\dot{a}}{a}\sum_{j\neq0}T_i^i=-\partial_t\rho c-3\frac{\dot{a}}{a}\rho c\\
\partial_{0}\sigma_{0}^{0}+\frac{3}{c}\frac{\dot{a}}{a}\sigma_{0}^{0}-\frac{1}{c}\frac{\dot{a}}{a}\sum_{j\neq0}\sigma_i^i=-K\partial_t\epsilon(t)
\end{gather*}
ce qui donne:
\begin{equation*}
\partial_t\rho(t)+3\frac{\dot{a}}{a}\rho(t)=-\frac{K}{c^2}\partial_t\epsilon(t) 
\end{equation*}
que l'on réécrit de façon plus élégante:
\begin{equation}\label{MFE2}
\partial_t(a(t)^3\rho(t))=-\frac{K}{c^2}a(t)^3\partial_t\epsilon(t) 
\end{equation}
On remarque que si $\rho(t)=\tilde{\rho}a(t)^{-3}$ où $\tilde{\rho}$ est une constante, alors pour que \eqref{MFE2} soit consistant nous devons avoir $\epsilon(t)=\varepsilon$, où $\varepsilon$ est une constante. On retrouve alors le modèle standard de la cosmologie. L'équation \eqref{MFE2} n'est pas nouvelle et a déjà été écrite dans des modèles supposant la constante cosmologique dépendante du temps. Cependant, nous établissons ici la correspondance suivante:
\begin{equation}
K\epsilon(t)=\frac{\Lambda(t) c^4}{8\pi G}\ , 
\end{equation}
c'est à dire que la fonction $\Lambda(t)$ est reliée au champ de déformation ainsi qu'au module d'élastricité isotrope. De ce point de vue on donne une interprétation à la constante cosmologique différente de l'interprétation habituelle qui l'associe à l'énergie du vide. 

En outre, nous avons établi d'après le principe (P3) et formalisé par les équations d'induction \eqref{E3}, que ce champ \textit{cosmologique} de déformation couple avec la matière en déformant les trajectoires géodésiques ou de façon équivalente, en générant un courant d'accélération non-nul. L'effet de la pression isostatique $-K$ a en quelque sorte pour effet de 'pousser' le gaz de matière. On rejoint cette interprétation en écrivant l'équation \eqref{E3} dans notre modèle cosmologique. Calculons d'abord les quantités suivantes:
\begin{gather*}
\epsilon_{0}^{\gamma}\partial_{\mu}T_{\gamma}^{\mu}=\epsilon_0^0\partial_0 T^0_0=\epsilon(t)\partial_t \rho(t)c
\Delta^{\mu}_{\mu\gamma}T_{0}^{\gamma}=\Delta^0_{00}T^0_0
\Delta^{\gamma}_{0\mu}T_{\gamma}^{\mu}=\Delta^0_{00}T^0_0
\end{gather*}
puisque $\epsilon_{\mu\nu}=0$ si $\mu$ ou $\mu$ $\neq0$. 
\footnote{Seule la composante $\epsilon_{00}=D_0 G_0=\epsilon(t)$ est non nulle car d'après le principe cosmologique les composantes $G_\mu$ ne peuvent dépendre que de la variable $t$ donc $\epsilon_{ij}=0,\ i,j\neq0$. De plus, le choix du référentiel des coordonnées comobiles nous permet de prendre nulles les composantes spatiales du vecteur de déformation $G_i=0,\ i\neq0$ puisque l'observateur en chaque instant est attaché à la matière en mouvement.} 
Par conséquent, d'après \eqref{E3} on a pour $\nu=0$: 
\begin{equation*}
\partial_t\rho(t)+3\frac{\dot{a}}{a}\rho(t)=\epsilon(t)\partial_t\rho(t) 
\end{equation*}
que l'on peut réécrire:
\begin{equation}\label{MFE3}
\partial_t(a(t)^3\rho(t))=a(t)^3\epsilon(t)\partial_t\rho(t)\ . 
\end{equation}
Pour résumer, nous avons les trois équations suivantes:
\begin{gather}
\frac{\dot{a}^2}{a^2}=\frac{8\pi G}{3c^2}\left(\rho(t)c^2+K\epsilon(t)\right)\label{I} \\
\partial_t(a(t)^3\rho(t))=-\frac{K}{c^2}a(t)^3\partial_t\epsilon(t)\label{II} \\
\partial_t(a(t)^3\rho(t))=a(t)^3\epsilon(t)\partial_t\rho(t) \label{III} 
\end{gather}
les deux groupes d'équations \eqref{II} et de \eqref{III} donnent:
\begin{equation}\label{IV}
-\frac{K}{c^2}\partial_t\epsilon(t)=\epsilon(t)\partial_t\rho(t) 
\end{equation}
qui nous permet d'exprimer l'amplitude des déformations $\epsilon(t)$ en fonction de la densité de matière:
\begin{equation*}
\epsilon(t)=A \exp{\left\{-\frac{c^2}{K}\int dt\partial_{t}\rho(t)\right\}}=A \exp{\left\{-\frac{\rho(t)c^2}{K}\right\}}\ . 
\end{equation*}
En notant $\varepsilon$ la valeur de l'amplitude à $t=t_0$ (i.e., à notre époque), on obtient:
\begin{equation}\label{S1}
\epsilon(t)=\varepsilon\cdot \exp{\left\{\frac{c^2}{K}(\rho(t_0)-\rho(t))\right\}}
\end{equation}
On trouve donc que l'amplitude des déformations varie en fonction du temps et dépend de la densité de matière dans l'espace.
Il est clair que pour des périodes $t>t_1$ où $t_1$ marque le début de l'époque où la matière domine, la densité diminue quand le temps augmente. Par ailleurs, on suppose, toujours pour $t>t_1$, que les déformations sont petites $\epsilon(t)\ll1$ et donc d'après l'équation \eqref{III} que $$\partial_t(a(t)^3\rho(t))\approx 0$$ ce qui donne en première approximation (à l'ordre zero des corrections) $$\rho(t)\approx\tilde{\rho}a(t)^{-3}$$ qui redonne la prédiction du modèle standard. Dans ce cas, d'après \eqref{S1} l'amplitude se comporte comme 
\begin{equation}\label{epsilon1erordre}
\epsilon(t)=\varepsilon\cdot \exp{\left\{\frac{\tilde{\rho}c^2}{K}a(t_0)^{-3}\left(1-\frac{a(t_0)^{3}}{a(t)^{3}}\right)\right\}}
\end{equation}
et étant donné que $a(t)$ croît à mesure que $t$ augmente, on a que $\epsilon(t)$ est une fonction croissante du temps. En outre, lorsque $t\rightarrow+\infty$, l'amplitude des déformations est constante:
$$\epsilon(+\infty)=\varepsilon\cdot\exp{\left\{\frac{c^2}{K}\rho(t_0)\right\}}$$
et on retrouve dans ce cas limite le modèle standard. 

Maintenant, on voudrait établir les corrections au premier ordre (i.e., en deuxième approximation) au modèle standard pour la densité de matière $\rho(t)$. D'après ce que nous venons de voir ci-dessus, l'amplitude $\epsilon(t)$ n'est pas constante. Mais on peut montrer que pour des petites déformations, la variation temporelle de $\epsilon(t)$ influence peu la densité de matière. On peut raisonnablement admettre dans un premier temps cette relation pour $t=t_0$:
\begin{equation}\label{KLambda}
K=\frac{\Lambda c^4}{8\pi G\varepsilon}\ , 
\end{equation}
où d'après les observations $\Lambda\sim 10^{-52}\ \mathrm{m}^{-2}$. Etant donné que $\varepsilon\ll1$, on trouve que $K\gg \frac{\Lambda c^4}{8\pi G}=\rho_\Lambda c^2$, où $\rho_\Lambda\sim 10^{-29}\mathrm{g.cm}^{-3}$ est la densité d'énergie du vide. Toujours d'après les observations, on sait que la densité d'énergie du vide est de l'ordre de celle de la matière (à un multiple près), on en conclut donc que le module d'élasticité isostatique $K$ est très grand devant la densité d'énergie de matière:
\begin{equation}\label{Kgrand}
K\gg \rho(t)c^2,\ t>t_1 
\end{equation}
\footnote{La théorie de déformation élastique considérée ici est linéaire. Cela ce trouve justifié pour le modèle cosmologique considéré ici car nous avons montré que de supposer les contraintes constantes pouvait être une bonne approximation pour l'ère matière dominante. En revanche, dans les périodes antécédentes il faudrait réécrire les équations et regarder comment évolue $\epsilon(t)$. Il est possible que dans ce cas $\epsilon(t)$ soit plus grand que 1. Si c'était le cas, la théorie linéaire ne serait plus valide. Il faudrait revenir sur cette hypothèse et formuler une théorie non-linéaire de déformation.} 
Ainsi en posant:
\begin{equation}\label{rho1erordre}
\rho(t)=\tilde{\rho}(1+\delta(t))a(t)^{-3},\ \delta(t)\ll1,\ \partial_t\delta(t)\ll\cdot{a}(t)/a(t)  
\end{equation}
et en isérant \eqref{rho1erordre} dans \eqref{III}, on trouve que
\begin{equation*}
\partial_t\delta(t)\approx a(t)^3\epsilon(t)\partial_t(a(t)^{-3})\approx\varepsilon a(t)\partial_t(a(t)^{-3})=-3\varepsilon\frac{\dot{a}(t)}{a(t)} 
\end{equation*}
où $\epsilon(t)\approx1$ puisque $\exp\left\{(\rho(t)c^2-\rho(t_0)c^2)/K\right\}\approx1$ d'après \eqref{Kgrand}.
On obtient donc:
\begin{equation}
\delta(t)\approx -3\epsilon\ln{\left\{\frac{a(t)}{a(t_0)}\right\}} 
\end{equation}
en fixant $\tilde{\rho}=\frac{a(t_0)^3}{\rho(t_0)}$.  

Il ne nous reste plus qu'à utiliser l'équation \eqref{MFE1} pour déterminer le facteur d'échelle en tenant compte des corrections \eqref{epsilon1erordre} et \eqref{rho1erordre}:
\begin{equation}\label{Cons1erordre}
\frac{\dot{a}^2}{a^2}\approx\frac{8\pi G}{3c^2}\left(\frac{\tilde{\rho}}{a(t)^3}\left(1+\delta(t)-\varepsilon\right)+K\varepsilon\left(1+\frac{\rho(t_0)}{K}\right)\right) 
\end{equation}
On voit donc qu'en deuxième approximation (corrections au premier ordre) on trouve une lois d'évolution similaire celle du le modèle standard mais qui dépend des paramètres $\epsilon$ et $K$ ainsi que d'un terme logarithmique supplémentaire. Notons qu'en prenant $t=t_0$ dans \eqref{Cons1erordre}, on retrouve la relation: $$1=\Omega_{M}+\Omega_{\Lambda}$$ avec $\Omega_M=\rho(t_0)/\rho_c$ et $\Omega_0=K\varepsilon/\rho_c$ (le produit $\kappa\varepsilon$ étant relié à $\Lambda$ par \eqref{KLambda}), où la densité critique est $\rho_c=3H_0 c^2/(8\pi G)$ et la constante d'Hubble est donnée par $\dot{a}(t_0)/a(t_0)$. 

Pour le moment nous nous contentons des corrections au premier ordre car elles fournissent des déviations au modèle standard qui semblent suffisantes étant donné que des observations actuelles \cite{CMB} sont déjà bien expliquées par ce modèle. Un travail d'analyse de données nous permettrait d'ajuster de façon optimale les constantes de notre modèle, à savoir $K$ et $\varepsilon$. Mais seules de futures observations plus précises montrant des déviations au modèle standard nous permettraient de conclure si la théorie proposée dans cet article est valide.

{99}

\end{document}